\documentclass[notitlepage,longbibliography]{revtex4-1}
\usepackage{graphicx}
\usepackage{hyperref}
\usepackage{amsmath}
\usepackage{color}

\begin{document}
\title{Probing chiral superconductivity in Sr$_{2}$RuO$_{4}$ \textit{underneath} the surface by point contact measurements}
\author{He WANG}
\author{Jiawei LUO}
\author{Weijian LOU}
\author{Jian WEI}\email{weijian6791@pku.edu.cn}
\affiliation{International Center for Quantum Materials, School of Physics, Peking University, Beijing 100871, China}
\affiliation{Collaborative Innovation Center of Quantum Matter, Beijing, China}

\author{J.E. Ortmann}
\author{Z.Q. Mao}
\affiliation{Department of Physics, Tulane University, New Orleans, Louisiana 70118, USA}

\author{Y. Liu}
\affiliation{ Department of Physics and Materials Research Institute, The Pennsylvania State University, University Park, Pennsylvania 16802, USA}
\affiliation{Department of Physics and Astronomy and Key Laboratory of Artificial Structures and Quantum Control (Ministry of Education), Shanghai Jiao Tong University, Shanghai 200240, China}
\date{\today}

\begin{abstract}
Sr$_{2}$RuO$_{4}$ (SRO) is the prime candidate for chiral $p$-wave superconductor with critical temperature $T_{c}(SRO)\sim$1.5 K.  Chiral domains with opposite chiralities $p_{x}\pm ip_{y}$ were proposed, but yet to be confirmed.  We measure the field dependence of the point contact (PC) resistance between a tungsten tip and the SRO-Ru eutectic crystal, where micrometer-sized Ru inclusions are embedded in SRO with atomic sharp interface. Ruthenium is an $s$-wave superconductor with $T_{c}(Ru)\sim$0.5 K, flux pinned near the Ru inclusions can suppress its  superconductivity as reflected from the PC resistance and spectra. This flux pinning effect is originated from SRO \textit{underneath} the surface and is very strong. To fully remove it, one has to thermal cycle the sample above $T_{c}(SRO)$. This resembles the thermal demagnetization for a ferromagnet, where ferromagnetic domains are randomized above its Curie temperature. Another way is by applying alternating fields with decreasing amplitude, resembling field demagnetization for the ferromagnet. The observed hysteresis in magnetoresistance can be explained by domain dynamics, providing support for the existence of chiral domains. The origin of strong pinning \textit{underneath} the surface is also discussed.

\end{abstract}
\maketitle

\section{Introduction}

The mechanism of the superconductivity (SC) of the layered perovskite ruthenate Sr$_{2}$RuO$_{4}$ (SRO), the prime candidate for topological chiral $p$-wave superconductivity, is not clear after over 30 years investigation~\cite{Maeno1994nature,Mackenzie2003rmp,Maeno2012jpsj,Kallin2012rpp,Liu2015pc}. Experimental results by muon spin-relaxation~\cite{Luke1998nature} and polar Kerr effect~\cite{Xia2006prl} suggest that the superconducting order parameter (OP) breaks the time-reversal symmetry and forms chiral domains with two different chirality ($p_{x}\pm ip_{y}$), similar to the domains in a ferromagnet. The existence of such chiral domains has not been conclusively confirmed, although there were indirect evidences: domain wall pinning was assumed to interpret the strong flux pinning (zero flux creep at lower temperatures) in \textit{bulk} magnetization relaxation measurements~\cite{Mota1999pb,Mota2000pc,Dumont2002prb}; and domain dynamics was also assumed to explain the field modulation of critical currents for corner junctions~\cite{Kidwingira2006science}. However, direct evidence of the chiral domains is absent. For example, edge current around domain walls and sample edges, which should lead to measurable magnetic fields, was not observed by \textit{local} field imaging methods on etched mesoscopic disks~\cite{Kirtley2007prb,Hicks2010prb,Curran2014prb,Lederer2014prb}, nor did early micro-Hall probe near the edge of SRO crystal~\cite{Tamegai2003pc}. 

Nevertheless, indirect evidence of the $p$-wave superconductivity was reported in local transport measurements, including (scanning) tunnel junction spectroscopy~\cite{Upward2002prb,Firmo2013prb,Kashiwaya2011prl,Kashiwaya2014pe} and point contact spectroscopy (PCS)~\cite{Laube2000prl,Wang2015prb}, where the order parameter symmetry may be inferred by fitting the conductance spectra~\cite{Kashiwaya2011prl,Kashiwaya2014pe}, but this method may not distinguish the chiral edge states from helical states~\cite{Scaffidi2014prb}. 

Different from the above-mentioned local transport measurements, here we conduct PC  measurements on SRO-Ru eutectic system~\cite{Maeno1998prl}. In the eutectic system, micrometer-sized Ru inclusions are embedded in SRO with atomic sharp interface. This well-defined interface, between an $s$-wave elemental superconductor and a presumed $p$-wave superconductor, may leads to a spontaneous flux distribution~\cite{Kaneyasu2010jpsj}, similar to the edge current at domain walls. The most interesting and unexpected outcome of the PC measurements, is that  point contact on Ru inclusions can be used for probing local flux: since Ruthenium is a superconductor with a lower critical temperature $T_{c}(Ru)\sim$0.5 K, its superconducting transition can be used to probe local flux. Compared with Hall effect sensor in field imaging~\cite{Kirtley2007prb,Hicks2010prb,Curran2014prb}, this approach is different in several aspects: (1) Ru inclusions are embedded in SRO matrix and largely \textit{underneath} the surface, so we can probe the flux without influence of the surface degradation; (2) Field imaging usually is done with field cooling and close-to equilibrium distribution of vortex, for PC measurements we focused on field sweeps and hysteresis; (3) The field range probed here is much larger than field imaging, which is usually in the low field limit due to the resolution of vortex. As will be shown, although the spontaneous flux is not fully confirmed, we observed strong flux pinning (MR hysteresis) and clear domain dynamics which are consistent with the chiral domain proposal. Additionally, the measured PC spectra may help to understand better about the interaction of the order parameters in two crystalline Ru and SRO with atomic sharp interface, which is an interesting topic by itself~\cite{Maeno1998prl,Yaguchi2003prb,Ying2009prl,Ying2013ncomms}.

\section{Experimental Methods}

\begin{figure*}
\includegraphics[width=16cm]{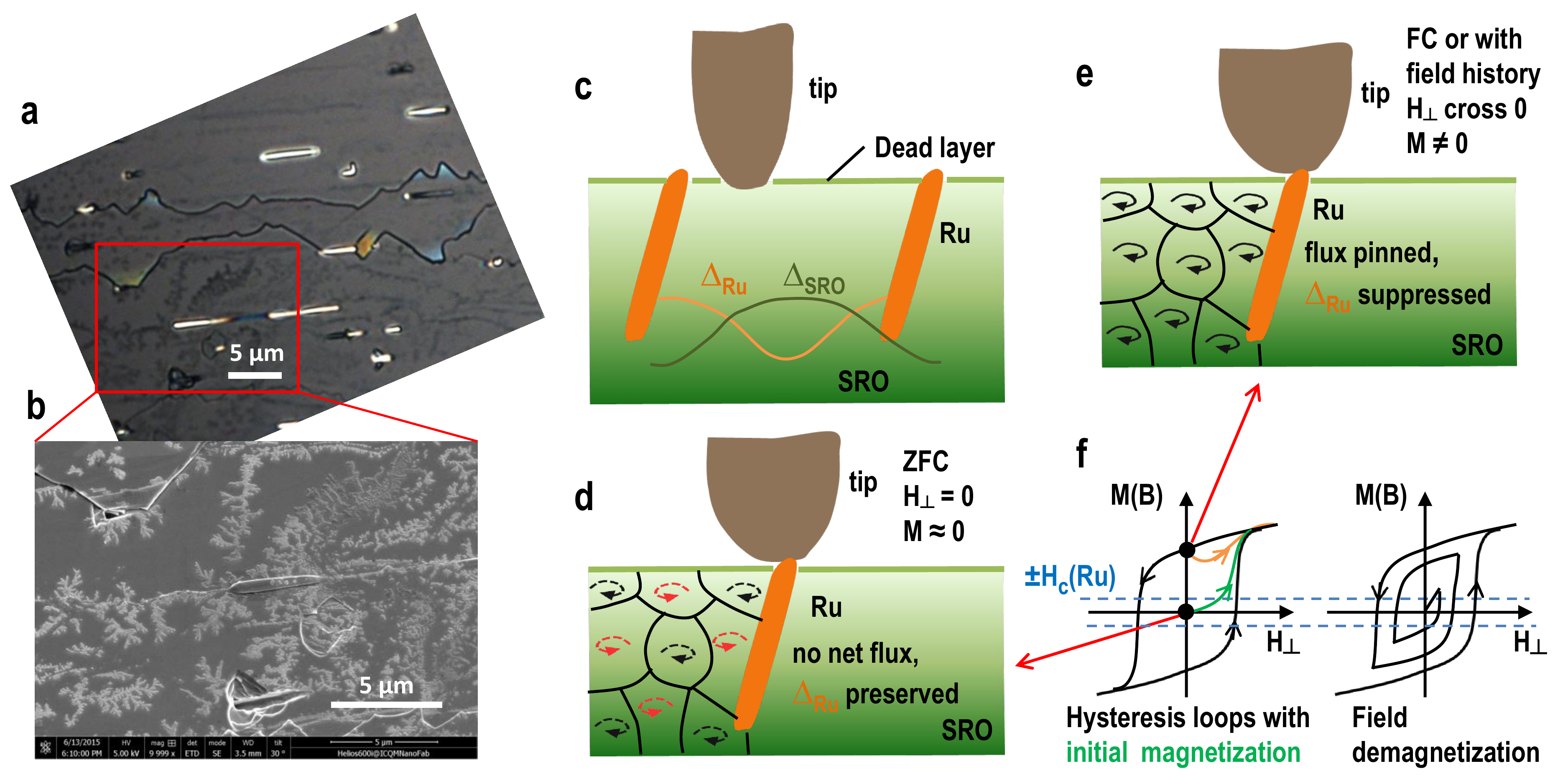}
\caption{Images of Ru inclusions on SRO surface and schematics for the point contacts (PCs). (a) Optical image showing Ru inclusions on the surface, which are about 1 $\mu$m wide with various lengths. The depth of these inclusions can be tens of  $\mu$m~\cite{Maeno1998prl} and is undetermined here. (b) SEM image of the region within the red box. The Ru inclusions clearly bulge from the SRO surface, and the dendritic patterns indicate surface degradation. The scale bar in both images is 5 $\mu$m.  After imaging the surface is scratched with a ceramic knife to expose a relatively fresh surface before the PC experiments.  (c) The schematic illustration of the PC between a tungsten tip and SRO surface (W/SRO-Ru), used to explain the two superconducting transitions shown in Fig.~\ref{fig_PC-1}a. Also shown is the assumed distance dependence of the superconducting order parameters of SC in Ru and SRO, without considering the coupling between them. The thick green line on the surface indicates the surface dead layer. (d) Blunted tip on a Ru inclusion (W/Ru-SRO). The chirality of the domains, $p_{x}\pm ip_{y}$, is shown by the clockwise or counter clockwise arrows, and is not fully developed for Zero Field Cooling (ZFC). With randomized chirality, the effective ``M'' is close to zero. (e) For field cooling, as well as for ZFC but with field history, the polarization of the chiral domains is induced and remains when the field is reduced to zero. The magnetization states in (d) and (e) are indicated by black dots in the hysteresis loops in (f). Also in (f), we show the initial magnetization curve (in green) which can be used for Fig.~\ref{fig_PC-2}(e), also show the reversal curve (in brown) when we restart the field sweep routine at zero field in reverse direction of the hysteresis loop, which can be used for Fig.~\ref{fig_PC-1}(e), and field demagnetization curve for Fig.~\ref{fig_PC-3}(d). The two horizontal dashed lines in blue denote the critical fields ($\pm H_{c}$) for Ru inclusions, within which the PC resistance shows a dip due to the recovered superconductivity. }  
\label{fig_schematic}
\end{figure*}

The rod-like Ru inclusions on the surface of the cleaved SRO crystal can be seen clearly in the optical microscope image Fig.~\ref{fig_schematic}(a), and more details are revealed in the SEM image Fig.~\ref{fig_schematic}(b), where the surface degradation can be observed for the crystal that was stored in ambient conditions. Before PC experiment, the surface is scratched with a ceramic knife to expose a relatively fresh surface. The SRO single crystals are grown by floating zone methods with Ru as self-flux~\cite{Mao2000mrb} with excess amount of Ru, so that an eutectic phase is formed with embedded Ru lamellar inclusions~\cite{Maeno1998prl}. The resulted SRO-Ru crystal has an extended superconducting transition critical temperature ($T_{c}$) from the intrinsic 1.5 K to about 3 K, and this $T_{c}$ enhancement is believed originated from the interface between Ru inclusions and SRO, possibly due to lattice distortions and strain although the exact mechanism is not confirmed~\cite{Maeno1998prl,Yaguchi2003prb,Ying2009prl,Ying2013ncomms}. In fact, it was found that uni-axial pressure on pure SRO can enhance SC as well~\cite{Kittaka2010prb,Hicks2014science,Wang2015prb,Steppke2016arXiv}. Regarding the symmetry of the OP for this enhanced SC, non-monotonic temperature dependence of the critical current (kinks near 1.5 K) and critical current switchings in so-called topological junctions suggested that the OP symmetry is different from that of pure  SRO~\cite{Nakamura2011prb,Anwar2013sr}. 

In Fig.~\ref{fig_schematic}(c) the schematic of the tungsten tip and the eutectic SRO-Ru crystal is shown. The point contact is made between a tungsten tip and the SRO crystal, which is fixed on a silicon chip and mounted on the attoCube nanopositioner stack. The tip and the nanopositioner stack are both secured on a metal housing which is suspended by springs to the end of a cold-insertable probe for a Leiden cryogen-free dilution fridge. The base temperature of the probe at the mixing chamber stage is about 0.1 K (12 mK without the point contact setup), but the base temperature for the sample stage is higher ($\sim$0.3 K) due to thermal loads from the wiring for the sample and nanopositioners, as well as Joule heating during the measurements. Differential resistance ($dV/dI$) is measured with standard lock-in technique with home-made battery-powered electronics to reduce external noises. For more details please refer to our previous study on pure SRO~\cite{Wang2015prb}.

There is always concern whether the surface property probed correlates with that of the bulk \textit{underneath} the surface. This is particularly critical for SRO since its surface undergoes reconstruction after cleaving and the SC may be destroyed, besides that surface contamination may also results in a dead layer. This might explain why the results of previous local transport measurements are not consistent with each other. One way to circumvent this problem is to make PCs with hard tips, thus the surface layer may be penetrated through by the tip~\cite{Gloos1996jltp_scaling,Gonnelli2002jpcs}. Using hard tungsten tips and our home-built point contact set-up,  we previously obtained reproducible result of the SC gap ($\sim$0.2 mV) for pure SRO~\cite{Wang2015prb}, consistent with that estimated by weak coupling theory, and different from the results from many other groups. 

Since Ru is much softer than the dead layer on SRO, for making PC on Ru inclusions it is not necessary to push the tip to penetrate the surface layer, as schematically plotted in Fig.~\ref{fig_schematic}(d). Nevertheless, in the beginning of the experiment, when the tip is sharp and can penetrate the surface dead layer, the interface can be made directly between the tungsten tip and the SRO, with Ru inclusions nearby (this type is referred as  W/SRO-Ru). In Fig.~\ref{fig_schematic}(c) we also plot the cross-sectional view of the PC and the distance dependence of the superconducting OPs for SC in Ru and SRO, without considering intermixing of the two OPs. Results for one of this kind PC showing both OPs, labelled as PC-1, are presented later in Fig.~\ref{fig_PC-1} due to the complexity.  Later in the experiment, the tip get flattened and the interface is normally between the flattened tip and the bulged Ru inclusions (W/Ru-SRO), as shown in Figs.~\ref{fig_schematic}(d) and \ref{fig_schematic}(e). Measurement results for two such PCs, labelled PC-2 (in the same run as PC-1) and PC-3 (in a later run), are presented in Figs.~\ref{fig_PC-2} and ~\ref{fig_PC-3} respectively. For PC-1 we focus on the hysteresis of magnetoresistance and order parameters shown in point contact spectra, for PC-2 and PC-3 we focus on thermal demagnetization and field demagnetization (see Fig.~\ref{fig_schematic}(f)).

In previous PC studies, only in a few cases vortex pinning in a conventional superconductor was considered~\cite{Shan2006prb,Martinez-Samper2000pc}. MR hysteresis observed here was not reported in previous PC measurements. It was usually assumed that the PC causes some damage and locally the SC is suppressed, thus vortices are trapped near the PC since the energy cost is lower. However, this is irrelevant since here for PC on Ru inclusions the origin of the magnetization (pinned flux) is not at the PC interface but within the SRO, thus not directly affected by any damage the point contact made on Ru inclusions. 

\section{Results}

For simplicity, we first describe the results for PC-2 and PC-3 where the tip get flattened and the interface is normally between the flattened tip and the bulged Ru inclusions (W/Ru-SRO).

\subsection{PC-2 and PC-3, on Ru inclusions}

\begin{figure*}
\includegraphics[width=16cm]{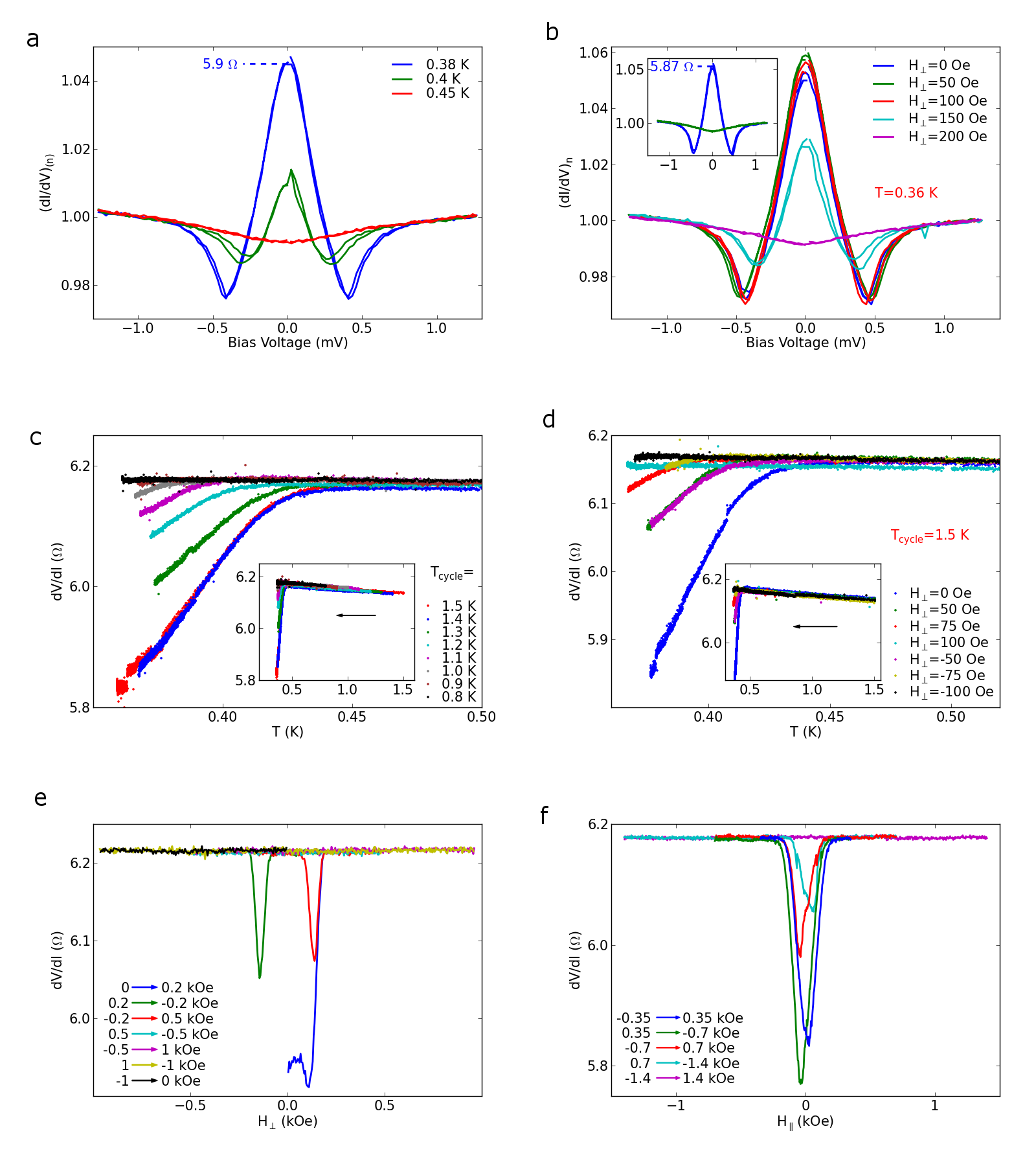}
\caption{Point contact measurements for PC-2 (W/Ru-SRO).(a) Normalized PC spectra at different temperatures from 0.38 to 0.45 K after ZFC. (b) Normalized PC spectra at 0.4 K under different $H_{\perp}$. Can see the quick change from 150 Oe. Inset shows the PC spectra without (blue) and with magnetic history (green) at T=0.38 K and zero field.  (c) Thermal demagnetization: after SC was suppressed by ramping $H_{\perp}$ at low temperature, warm up at zero field to different cycling temperatures $T_{cycle}$, and then zero field cool-down.  $T_{c}$ is shifted with different $T_{cycle}$. Inset: $R(T)$ curves up to 1.5 K. (d) $R(T)$ curves during field cool from 1.5 K with opposite polarities. Inset: $R(T)$ curves up to 1.5 K. (e) MR curves for $H_{\perp}$  at T=0.38 K. (f) MR curves for $H_{||}$  at T=0.38 K.  }  
\label{fig_PC-2}
\end{figure*}

\begin{figure*}
\includegraphics[width=16cm]{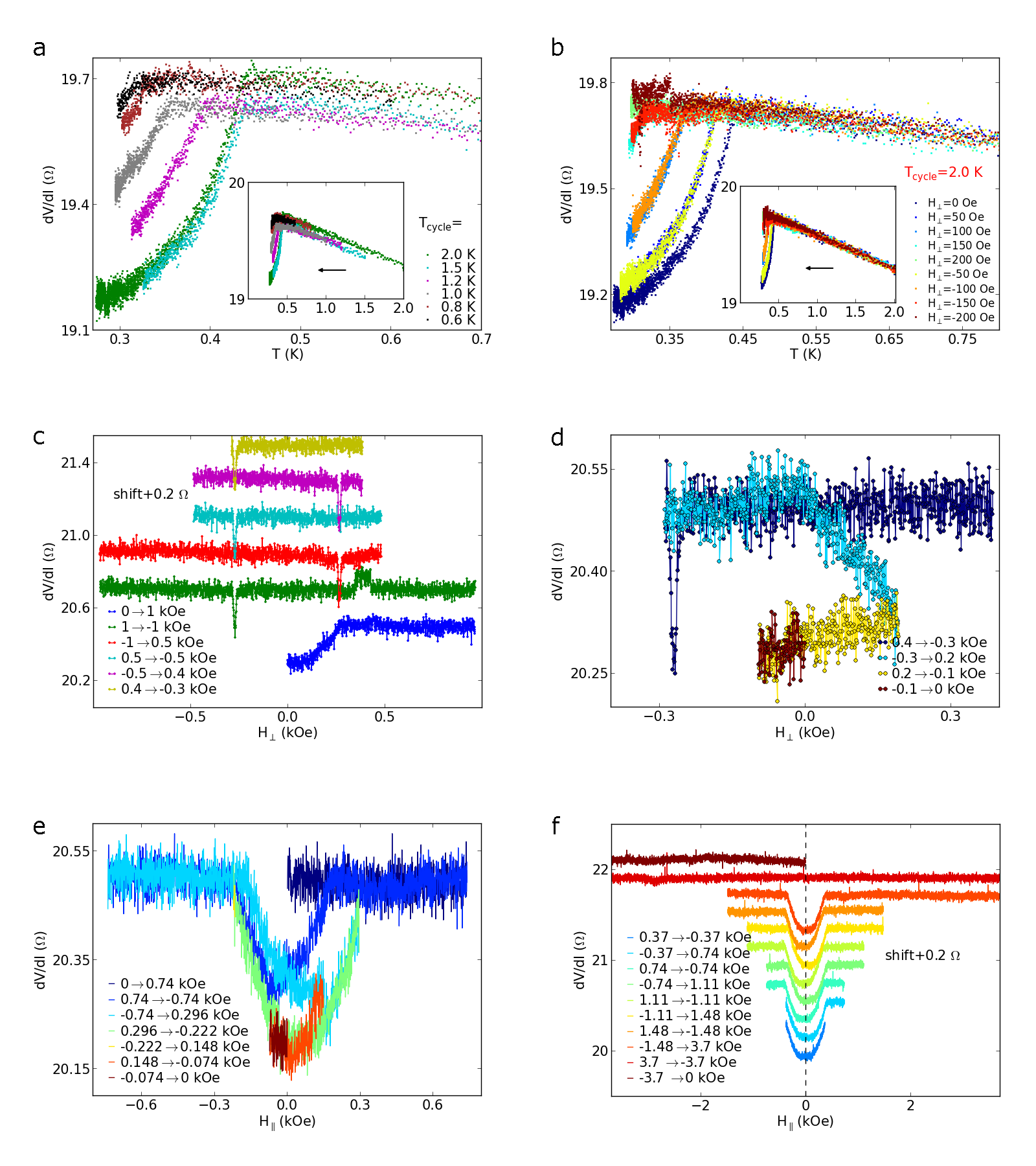}
\caption{Point contact measurements for PC-3 (W/Ru-SRO).(a) Thermal demagnetization: after SC was suppressed by ramping $H_{\perp}$ at low temperature, warm up at zero field to different cycling temperatures $T_{cycle}$, and then zero field cool-down.  Inset: $R(T)$  curves up to 2 K.  (b) $R(T)$ during cool-down at different fields. Inset: $R(T)$  curves up to 2 K.  (c) MR for $H_{\perp}$ at T=0.35 K showing resistance dips at around $\pm$0.27 kOe,  depending on the field sweeping direction. (d) Field demagnetization by continuing the field sweep with decreasing field amplitude (lower than the coercive field 0.27 kOe). The SC in Ru inclusion is partially recovered and the lowest resistance at zero field is comparable to that at the dip position. (e) MR for $H_{||}$ at 0.38 K showing a similar field demagnetization effect after the SC was totally suppressed initially. (f) MR for $H_{||}$ with increasing field amplitude at 0.35 K, after zero field cool. After the field is ramped to 3.7 kOe, the superconducting state at zero field is suppressed. For clarity a shift of +0.2  $\Omega$  has been done for consecutive curves in (c) and (f).  }
\label{fig_PC-3}
\end{figure*}

During zero field cooling, as shown in Figs.~\ref{fig_PC-2}(c) and \ref{fig_PC-3}(a), for PC-2 (PC-3) there is only one quick drop around 0.44 K (0.42 K), which is originated by the superconductivity in Ru. The deviation from $T_{c}(Ru)\sim$0.5 K could be due to inaccurate temperature measurement: the thermometer was put on the same substrate with the sample nearby but not in direct contact, also the local temperature at the point contact maybe different from that of the substrate.  

For PC-2, the conductance enhancement at small bias is suppressed by raising temperature above 0.5 K or by ramping field above 200 Oe (Figs.~\ref{fig_PC-2}(a) and \ref{fig_PC-2}(b)), consistent with that this conductance enhancement is indeed sustained by SC in Ru inclusions. Note that after the field is ramped back from 200 Oe back to zero, the conductance enhancement is still suppressed, as shown in the inset of Fig.~\ref{fig_PC-2}(b), suggesting remnant flux (see Figs.~\ref{fig_schematic}(e) and \ref{fig_schematic}(f)). This hysteresis behaviour is better illustrated by the magnetoresistance (MR) at zero bias, as shown in Fig.~\ref{fig_PC-2}(e).  This hysteresis reminds us the magnetization curve of a ferromagnet, e.g., see the textbook~\cite{Cullity2009} (especially chapter 16 where the same terminology can be used for ferromagnetism and superconductivity). For simplicity of the model, we assume a soft ferromagnet with reasonable large susceptibility, thus $\mathbf{M}\approx\mathbf{B}$, and at the coercive field $\mathbf{H_{coer}}\sim$ 200 Oe, both $\mathbf{M}$ and $\mathbf{B}$ come across zero. Similar $\mathbf{H_{coer}}$ were observed for another two PCs in the same run, see Supplementary Fig.~S2, and also for PC-1 in Fig.~\ref{fig_PC-1}(e). 

When started at zero field after ZFC, there is a small drop of resistance at around 100 Oe and then a quick increase to the normal state at field close to 200 Oe, indicating first enhancement and then suppression of SC in Ru. The initial enhancement with increasing field could be due to some proximity effect from SRO which may reduce the flux in the Ru (Fig.~\ref{fig_schematic}(d)), or possibly related to the spontaneous flux proposed at the interface between an $s$-wave superconductor and a $p$-wave superconductor~\cite{Kaneyasu2010jpsj}. By further increasing field close to $\mathbf{H_{coer}}$, the chiral domains get polarized since each chirality prefers certain polarity of the field, and the resulted large $\mathbf{M}$ leads to total suppression of SC in Ru inclusions, even after the field is reduced to zero (Fig.~\ref{fig_schematic}(e)).  

With small field sweeping amplitude the resistance dips near $\pm\mathbf{H_{coer}}$ are observed, as shown in Fig.~\ref{fig_PC-2}(e) by the  the green line from 0.2 to -0.2 kOe, and the red line from -0.2 to 0.5 kOe. These dips imply sign changing of the chiral polarization.   But with increasing field sweeping amplitude, the resistance dips were not observed for PC-2. This is probably because the growth of chiral domains is affected by the rate and amplitude of sweeping field, and the bigger domains (or stronger pinning force) induced by higher fields cause the switching of polarization too fast to resolve (the magnetization curve is still there but the probe is too slow to follow). Disappearing of the dip was also observed for another two PCs in the same run, see Supplementary Fig.~S2. Above $\mathbf{H_{coer}}$ there is almost no resistance change (in Fig.~\ref{fig_PC-2}e), indicating no influence of SRO on PC resistance.  

\textbf{Resemblance to thermal demagnetization.} It was proposed that chiral domains behave like ferromagnetic domains, thus it is natural to consider how the polarization (magnetization) of the chiral (ferromagnetic) domains may be randomized. The magnetization of a ferromagnet can be demagnetized by thermal or cyclic field methods~\cite{Cullity2009}. For thermal demagnetization, the ferromagnetic domains are randomized after the sample is heated above its Curie point and cooled in the absence of field. Similarly, the suppressed SC at lower temperatures (due to large $\mathbf{M}$, see Figs.~\ref{fig_schematic}(e) and \ref{fig_schematic}(f)) can indeed be recovered by thermal cycling to $T_{c}(SRO)\sim$1.5 K and then cool down in the absence of field. As show in Fig.~\ref{fig_PC-2}(c), when the thermal cycling temperature $T_{cycle}$ is lower than $T_{c}(SRO)$, the SC in Ru cannot be fully recovered and $T_{c}(Ru)$ is still suppressed.  This proves that the suppression of SC in Ru is due to remnant magnetization (chiral domain polarization) in SRO. The $T_{cycle}$ needs not reach the enhanced $T_{c}$ for the 3-K phase for the following reasons: first the volume of the 3-K phase with enhanced $T_{c}$ is small, thus has a negligible influence on the PC; secondly the OP of the 3-K phase may not be chiral $p$-wave~\cite{Maeno1998prl} so there is no intrinsic vortex pinning mechanism.  

This flux pinning effect is very symmetric to external fields. Starting from the same $T_{cycle}$ of 1.5 K,  cooling with opposite fields ($\pm$ 50 Oe, $\pm$ 75 Oe, $\pm$ 100 Oe etc, as shown in Fig.~\ref{fig_PC-2}(d)), gives similar $R(T)$ curves. These FC curves are also similar to those with different $T_{cycle}$'s (compare Fig.~\ref{fig_PC-2}(c)). Such resemblance corroborate with the existence of remnant magnetization when $T_{cycle}<$1.5 K. 

\textbf{Resemblance to field demagnetization.} The other option to revert to the magnetic ``virgin'' state is to follow the field demagnetization procedure for the ferromagnet (see Fig.~\ref{fig_schematic}(f)), by applying alternating fields of decreasing amplitude, which makes the domains smaller and/or randomly aligned. For PC-2, we only tried with increasing amplitude of sweeping fields. 
Later in another run with the same crystal, we made PCs on Ru inclusion similar to PC-2, and label one of them as PC-3. We start from large alternative fields and then reduce the amplitude, also we minimize the waiting time between the measurements of two consecutive data points, and reduced the step of field ramping, so the resistance drop can not be too fast to be registered. Then the resistance dips can be repeatedly observed, as shown in Fig.~\ref{fig_PC-3}(c).  The dip position now changed to around $\pm$270 Oe, higher than $\pm$200 Oe for PC-1 and PC-2, and the width of the dip is much narrower, suggesting faster domain dynamics (narrow resistance dips were also reported for magnetic insulator reflecting magnetic domain dynamics~\cite{Ma2015science}). Moreover, when the amplitude is further decreased from 0.4 $\rightarrow$ -0.3 $\rightarrow$ 0.2 $\rightarrow$ -0.1 $\rightarrow$ 0 kOe, as shown in Fig.~\ref{fig_PC-3}(d), the resistance decreases and reaches a similar value as the bottom of the resistance dip at $\mathbf{H_{coer}}$, suggesting the local remnant field $\mathbf{M}$ is minimized.  

For comparison, thermal demagnetizaton and field cool with different fields are also shown for PC-3 in Figs.~\ref{fig_PC-3}(a) and \ref{fig_PC-3}(b) respectively. Similar to that for PC-2 (Figs.~\ref{fig_PC-2}(c) and \ref{fig_PC-2}(d)), here the remnant magnetization can be completely removed after thermal cycle to 1.5 K, and for field cool there is excellent symmetry for both field polarities.


\textbf{Parallel field MR.} The parallel field MR data are less understood, but still are briefly presented here for completeness (additional results of tilted field and in-plane anisotropic field MR were not reported). SRO is a layered superconductor with much different in-plane $H_{c2|| c}(0)\sim$1.50 T and out-of-plane $H_{c2\perp c}(0)\sim$0.075 T~\cite{Mackenzie2003rmp}, thus $H_{||}$ with amplitude comparable to $H_{c2\perp c}(0)$ is not expected to affect SC in the SRO.   However, for PC-2 on Ru inclusions (W/Ru-SRO) we do observe resistance dips in the parallel field MR (see Fig.~\ref{fig_PC-2}(f)), with even smaller $\mathbf{H_{coer}}$, and the dip is also broader, about 200 Oe. We note that this is larger than $H_{c}$ of elemental superconductor Ru, which is about  25 Oe at 0.4 K (70 Oe at zero temperature)~\cite{Geballe1961prl,Gibson1966pr}. Another difference compared with perpendicular field MR, is that while with smaller sweeping field amplitudes the resistance dip appears in the opposite side after sweeping across zero field (to depolarize the remnant field at $\mathbf{H_{coer}}$), with larger sweeping fields the resistance dip appears in the same side towards zero field, e.g., -0.7 $\rightarrow$ 0.7 kOe the dip appears near -40 Oe, and 0.7 $\rightarrow$ -1.4 kOe, the dip at around 60 Oe.


Parallel field MR for PC-3 are also shown in Figs.~\ref{fig_PC-3}(e) and \ref{fig_PC-3}(f) for decreasing and increasing amplitude respectively. The resistance minimum near zero field gets lower for consecutive sweeps with decreasing amplitude, bearing some similarity to the MR with  $H_{\perp}$ (Fig.~\ref{fig_PC-3}(d)). For field sweeping with increasing amplitude, similar to that for PC-2 (Fig.~\ref{fig_PC-2}(f)), while at smaller field amplitudes the resistance minimum appears in the opposite side after crossing zero field , at larger amplitude the resistance minimum moves towards zero field. Additionally, for field cool, $H_{||}$ has a similar but smaller suppression effect on SC of Ru inclusions(as shown in Supplementary Fig. S3). One may think that there is finite $H_{\perp}$ component at the interface due to the inclined Ru/SRO interface, since the Ru inclusions are not aligned with any crystal orientation. However, this is not consistent with the smaller $\mathbf{H_{coer}}$ observed.

\subsection{PC-1, near the SRO-Ru interface}

\begin{figure*}
\includegraphics[width=16cm]{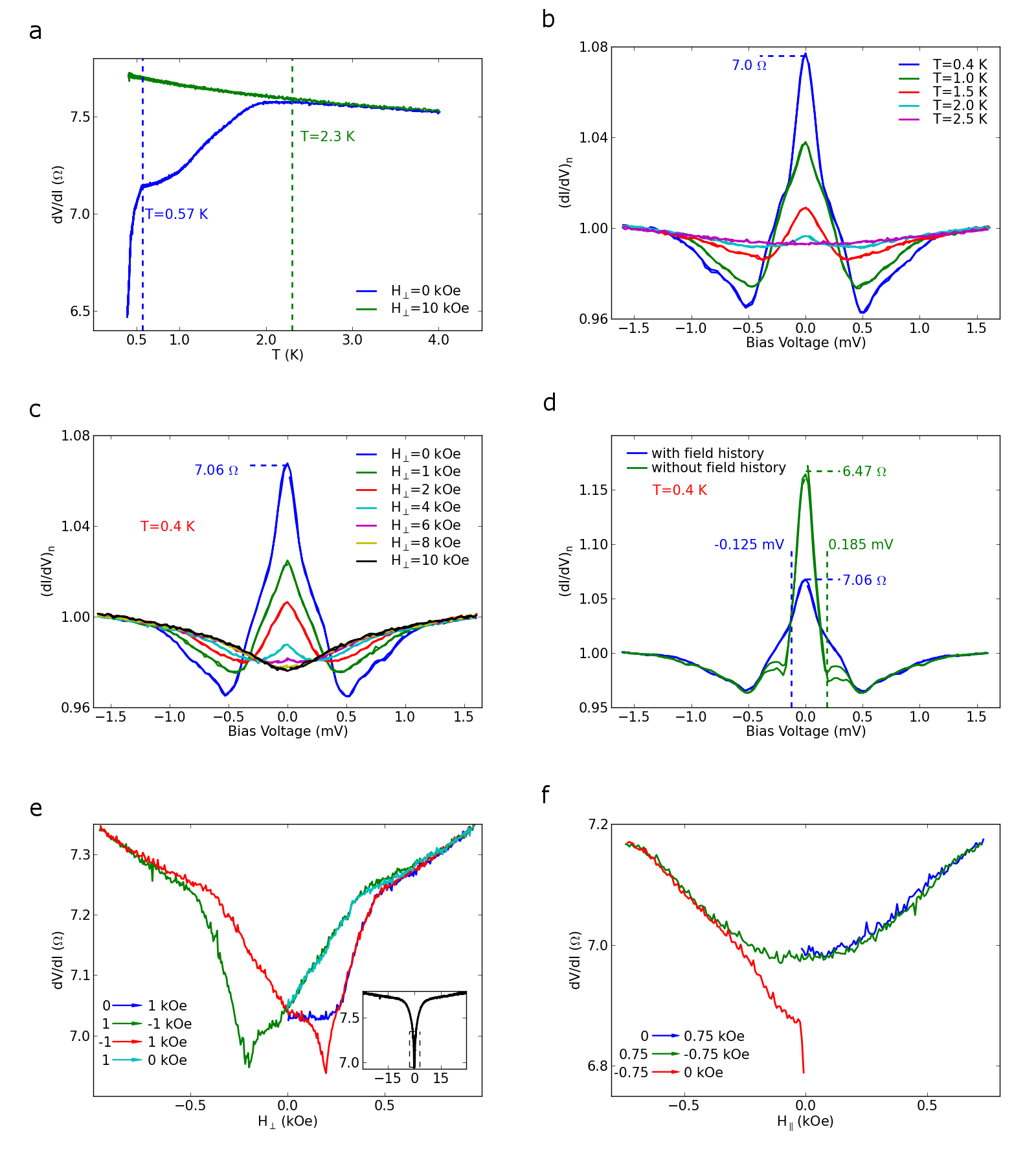}
\caption{Results for PC-1 (W/SRO-Ru). (a) Temperature dependence of the  resistance in zero field cooling process shows two drops indicating the superconducting transitions of the SRO eutectic phase (from 2.3 K) and Ru inclusion (from 0.57 K)  respectively. The transitions are totally suppressed with a perpendicular (out-of-plane) magnetic field of 10 kOe. Normalized differential conductance with field ramping history (b) at different temperatures from 0.4 to 2.5 K at zero field, (c) at 0.4 K under different perpendicular magnetic fields ($H_{\perp}$), and (d) compared with that without field history.  The highest conductance at zero bias (lowest resistance) is marked in (b)-(d). (e) Hysteresis of the magnetoresistance (MR) at T=0.4 K.  Inset: MR with a larger sweeping amplitude, the dashed box denotes the field range in the main panel. (f) MR with in-plane field ($H_{||}$) at T=0.4 K. In (e) and (f) different colors of the curves indicate the consecutive order of the field ramping. }  
\label{fig_PC-1}
\end{figure*}

For PC-1, the sharp tip penetrates through the dead layer (see Fig.~\ref{fig_schematic}(c)), so SC in both SRO and Ru can be clearly probed. The SC in SRO is shown by the gradual resistance drop starting from about 2.3 K to 0.57 K (Figs.~\ref{fig_PC-1}(a)), and the quick drop below 0.57 K is due to SC of Ru. Note that the SRO/Ru interface does not contribute to the PC resistance, since the interface resistance is usually in m$\Omega$ range (the interface area is usually on the order of 10 $\mu m^{2}$~\cite{Nakamura2011prb,Anwar2013sr}), much smaller than the PC resistance. 

With two OPs involved, we expect to see something different in the point contact spectra at finite bias. As can be inferred by the temperature and field dependence of the point contact spectra (Figs.~\ref{fig_PC-1}(b)-(d)), the conductance dips at around $\pm$0.5 mV  reveal information of the OP of SRO (see supplementary Figs.S4 and S5 for similar feature for PC-2 and PC-3). The $\pm$0.5 mV conductance dips and the broad zero bias conductance hump are suppressed when temperature is raised to 2.5 K or the out-of-plane field ($H_{\perp}$) increased to 10 kOe. Surprisingly, for  PC-2, similar feature is suppressed at 0.45 K and 200 Oe. This suggests that for W/Ru-SRO type contacts, while the PC spectra are influenced by SC in SRO (large gap value), it is still sustained by the conventional SC in Ru. Note that the gap value for element Ru is about 0.07 mV using the mean field estimation with $T_{c}$=0.5 K, much smaller than 0.5 mV observed here. 

The most noticeable difference compared with the spectra of PC-2 is an additional deflecting point at around $\pm$0.125 mV,  as marked by the blue vertical dashed line, for the blue curve in Fig.~\ref{fig_PC-1}(d). This feature is observed for the PC spectra of PC-1 at zero field with field history (finite $\mathbf{M}$). When there is no field history (zero field cooling, or ZFC), the spectra is different. As shown by the green curve in Fig.~\ref{fig_PC-1}(d), it evolves to an additional conductance dip at $\pm$0.185 mV,  as marked by the green vertical dashed line. The highest conductance without field history (6.47 $\Omega$) is close to that shown in the ZFC $R(T)$ curve (6.5 $\Omega$, Fig.~\ref{fig_PC-1}a), as this differential conductance was measured right after ZFC. After field ramping, the highest  conductance is reduced (7.06 $\Omega$), and the double dips at around $\pm$0.185 mV and $\pm$0.5 mV reduce to dips at $\pm$0.5 mV only, indicating OP in Ru is suppressed. So this is consistent with that Ru inclusions serve as local probe for $\mathbf{M}$.

\textbf{Hysteresis in magnetoresistance.}  For PC-1 there is clearly a MR  hysteresis in the  perpendicular applied field (Fig.~\ref{fig_PC-1}(e)),  similar to what we observed previously for PC on pure SRO~\cite{Wang2015prb}. The difference is that here there are additional resistance dips at $\mathbf{H_{coer}}\pm$200 Oe instead of a rounded valley (see Supplementary Fig.~S1 for comparison, for pure SRO, there are even Barkhausen-type jumps similar to the real magnetic domain dynamics). Such similar hysteresis confirms that the strong flux pinning is not due to the additional Ru inclusions but originated from SRO itself. Also this MR hysteresis is not an artifact as the hysteresis is not observed in PC measurements for other superconducting materials with similar set-up. 

The presence of the resistance dip depends on the field history. For example, if reverse the sweeping direction at zero field, as can be inferred from the hysteresis loops in Fig.~\ref{fig_schematic}(f), then there is no $\mathbf{M}$=0 point in that direction until sweep back again and cross zero.  As shown in Fig.~\ref{fig_PC-1}e, the field sweep follows the same routine, 0 $\rightarrow$ 1 $\rightarrow$ -1 $\rightarrow$ 1 $\rightarrow$ 0. So the last sweep is 1 $\rightarrow$ 0 kOe. Now we restart the field sweep at zero field,   0 $\rightarrow$ 1 kOe, opposite to the last sweeping direction, then there is no resistance dip at 200 Oe. When we sweep downward, 1 $\rightarrow$ -1 kOe, the resistance dip appears at -200 Oe, both consistent to our model.  

The lowest resistance at the resistance dip position is around 6.9 $\Omega$, still larger than the resistance observed during ZFC (6.5 $\Omega$, see Fig.~\ref{fig_PC-1}a). This may be understood if the suppression of SC in Ru is only partially reduced. One possible scenario is that: for multiple domains although the total $\mathbf{M}$ is zero at $\mathbf{H_{coer}}$, but locally there could be inhomogeneous flux and $\mathbf{M}$ is non-zero. For ZFC, domains were not magnetized (or not trained, as indicated by the dashed arrows in Fig.~\ref{fig_schematic}(d)) so there is no net flux.  

For parallel field MR, as shown in Fig.~\ref{fig_PC-1}(f), the data are less understood as in the case of PC-2 and PC-3. There is no hysteresis, nor  resistance dips at $\mathbf{H_{coer}}$. Instead, the resistance shows a broad minimum near zero field, and it is even possible to induce a sharp resistance drop near zero field. The broad minimum is consistent with the field dependence of a conventional superconductor. But if compared with the resistance in the perpendicular field MR (Fig.~\ref{fig_PC-1}(f)), this resistance change mostly corresponds to that for SRO, and only the sudden resistance drop may be related to the Ru inclusion. In the scenario of domain dynamics, this may be interpreted as that the chiral domains with out-of-plane polarization are randomized by $H_{||}$. 

\section{Discussions}

Clearly there is strong vortex pinning in SRO, but its origin is still not certain. There are a few theoretical proposals available to understand the vortex state in SRO. First, Sigrist and Agterberg studied the role of chiral domain walls on the vortex creep dynamics~\cite{Sigrist1999ptp}, which was used to explain the zero flux creep observed by Mota group~\cite{Mota1999pb,Mota2000pc,Dumont2002prb}. In this picture the domain walls are pinned at impurities and lattice defects so they do not move easily (this is how the domain picture in Fig.~\ref{fig_schematic}(d) and \ref{fig_schematic}(e) is derived).  Second, Garaud et al~\cite{Garaud2012prb} consider SRO as a  type-1.5  superconductor  with  long-range attractive, short-range repulsive intervortex interaction. This is used to explain the  vortex  coalescence observed by scanning Hall probe microscopy~\cite{Curran2014prb} and  possible clusters of vortices nucleating within a Meissner-like state implied by the $\mu$SR measurements~\cite{Ray2014prb}. Third, Ichioka et al~\cite{Ichioka2005prb} used the  time-dependent  Ginzburg-Landau  theory to study the magnetization process and found that with increasing magnetic fields, the domain walls move so that the unstable domains shrink to vanish, and the single domain structure is realized at higher fields. Along this line there were theories of doubly quantized vortices and other exotic behaviours that may lead to a nonzero chirality degeneracy broken field~\cite{Garaud2016prb,Sauls2009njp}.  Note that compared to the first proposal, chiral domain walls pinning is not emphasized, which to some extent suggest pinning by domain themselves and probably more relevant to our observations here. 

The chiral domain wall pinning proposal, was developed to understand the systematic experimental results of \textit{bulk} magnetization relaxation by Mota group~\cite{Mota1999pb,Mota2000pc,Dumont2002prb}, where a novel strong flux pinning (even zero flux creep at the lowest temperatures) was found and the higher the cycling magnetic field, the stronger the pinning effect. This is considered as indirect evidence of chiral domains. In our work the MR hysteresis also suggests strong pinning, but there are a few differences: 1) Previously the chiral domain wall pinning scenario seems inconsistent with the cycling field effect, since with higher cycling field, ``polarization'' of the chiral domains in SRO is enhanced and domain walls are reduced, resulting in less pinning. Here we propose that the domains themselves can provide strong pinning (once they are formed with field history), and compare the chiral domain dynamics with that of a ferromagnet,  by assuming  a local ``magnetization'' ($\mathbf{M}$) due to chirality polarization, the $\mathbf{H_{coer}}$ is a natural explanation for the necessity of a high cycling field. 2) Previously the relaxation measurement can not be done at zero field, but here we can measure the point contact spectra in the ZFC situation, which probes the ``virgin'' state without magnetization. 3) Previously the focus was the zero flux creep regime at the lowest temperatures (50 mK and lower), while here the focus is the domain dynamics at higher temperatures (but still much lower than $T_{c}(SRO)$). 4) Previously, strong pinning for both $H_{||ab}$ and $H_{\perp ab}$ is observed , here only strong hysteresis for $H_{\perp ab}$ is observed. The chiral domain walls should only exist in the $ab$ plane if the 2-$d$ $\gamma$ band is the active superconducting band, thus it is not clear whether the strong pinning for $H_{||ab}$ is due to the same mechanism. Further experiment, e.g., ac susceptibility study on the in plane metastable vortex state~\cite{Shibata2015prb}, maybe helpful to investigate this issue.


The second proposed model to understand vortex clustering on SRO surface is type-1.5  superconductivity, which was first named after the observation of vortex clustering on the surface of MgB$_{2}$~\cite{Babaev2005prb,Moshchalkov2009prl}, a two-band superconductor that has two weakly coupled order parameters with $\kappa_{1} < 1/ \sqrt{2}$ and $\kappa_{2} > 1/\sqrt{2}$. In fact, for single band superconductors with $\kappa \sim 1/\sqrt{2}$, there was also such long range attractive force~\cite{Auer1973prb}. For SRO, the in plane $\kappa_{ab}$=2.3 and out of plane $\kappa_{c}$=46, both are in type II regime~\cite{Mackenzie2003rmp}. So to apply this theory model, Garaud et al~\cite{Garaud2012prb} assume there are several coherence lengths in multicomponent superconductors~\cite{Garaud2012prb}, and find that  type-1.5 behavior can occur in multiband chiral Ginzburg-Landau theories for SRO. This may explain the clustering of vortex imaged by scanning field probe at low fields~\cite{Curran2014prb, Hicks2010prb} and bulk Meissner-like state implied by the $\mu$SR measurements~\cite{Ray2014prb}. However, simply type 1.5 superconductivity can not explain the observed MR hysteresis here for several reasons: 1) Similar MR hysteresis has not been found in the point contact measurements for MgB$_{2}$~\cite{Daghero2010sst}. 2) The difference between types 1.5 and 2 is usually for low field region, but here the temperature and field range are outside the typical regions for Ginzburg-Landau theories. The dynamics here may involve only fully penetrated vortex domains instead of vortex domains mixed with Meissner-like domains in the low field regime.3) For a system close to type I, one do not expect to see hysteresis in M(H).  

So let consider the last model, i.e., pinning by chiral domains themselves.  This model can explain the striking similarity to the ferromagnetism here. After ZFC, the applied field leads to the formation of chiral domains similar to ferromagnetic domains, which themselves become high energy barriers for flux, instead of resorting to domain wall pinning.  This is also consistent with the $\mu$SR experiment, that a large fraction of the volume is vortex free until the field ramped to above 100 Oe. By assigning the quite large observed $\mathbf{H_{coer}}$ as the flip field for chiral domains, we have to abandon the  previous belief that chiral domains flips easily~\cite{Tamegai2003pc,Curran2011prb}. In fact, $\mathbf{H_{coer}}$ is  much larger than $H_{c1}$ measured by local magnetization hysteresis loops with Hall probe~\cite{Tamegai2003pc}, and close to the thermal dynamic critical field~\cite{Mackenzie2003rmp}.

We note that, the proposed chiral domains should not be mixed with conventional vortices domains, since it is possible to push vortices of different vorticity into the chiral domains that with a preferred vorticity, jut with different energy cost. And the strong vortex pinning by chiral domains is absent for conventional vortex domain pinning by domain walls.
From the MR results here, there is another feature probably pointing to unconventional vortex pinning, i.e., for regular vortex pinning the MR usually does not show exact symmetry respect to zero field~\cite{Kidwingira2006science}, but here it does.  There was a concern for the chiral domain wall pinning scenario that domain wall pinning was also suggested for UPt$_3$~\cite{Dumont2002prb}, but in later experiments it seems that single domain without domain walls was inferred~\cite{Strand2010science,Schemm2014science}. If pinning is by domains themselves, not by domain walls, then there is no more concern. There was also a disparity regarding chiral domain size~\cite{Kallin2009jpcm}, which was estimated to be around 100 $\mu$m or larger in the polar Kerr effect measurement~\cite{Xia2006prl}, and about 1 $\mu$m in the critical current measurements for corner junctions~\cite{Kidwingira2006science}. This can be reconciled if the domain size can be determined by the internal defects which are very sample specific due to sample growth parameters, and then it may also be determined by the alignment in a multiple domain assembly, as draw in the schematic illustration in Fig.~\ref{fig_schematic}.  

As an additional note, by the MR hysteresis along, it seems difficult to distinguish strong flux pinning effect from the possible coexistence of ferromagnetism and superconductivity, the latter of which was proposed for the interface superconductivity between LAO/STO~\cite{Dikin2011prl}, also an intriguing subject. From the aspect of time reversal symmetry breaking, the difference between ferromagnetism and equal-spin triplet pairing, probably is that the former has a static ferromagnetic order parameter, while the latter does not.

\section{Conclusion}

By using the Ru lamella inclusions embedded in single crystal SRO as local magnetization sensor, we found a new method to probe the local flux underneath the surface. The observed strong MR hysteresis with applied field perpendicular to the $ab$ plane, and various field dependences (thermal demagnetization, field demagnetization, coercive field etc) indicate striking similarity to ferromagnetic domains with one easy axis. Such similarity provides indirect evidence of chiral domains and domain dynamics. We also discussed possible intrinsic pinning mechanisms, including the chiral domain wall pinning~\cite{Sigrist1999ptp} and the type-1.5  superconductivity~\cite{Babaev2005prb,Moshchalkov2009prl}, but both seem to have difficulties to explain the hysteresis. And the left explanation is pinning by chiral domains themselves. Besides the domain dynamics obtained by zero bias point contact resistance, the point contact spectra at finite bias manifest the order parameters of both Ru and SRO, thus might be helpful for understanding the interaction between them, although more work needs to be done to clarify this. Further experimental investigation along this direction includes possibly scanning point contact measurement to check the proximity effect near Ru inclusions, and PC with ferromagnetic or s-wave superconducting tips.

\section*{Acknowledgements}
Jian Wei thanks Yoshiteru Maeno, Venkat Chandrasekhar, Jim Sauls,  William P. Halperin, Zhili Xiao, Haihu Wen, Peng Xiong, and Laura Greene for  discussions, and especially Weida Wu for the suggestion of field demagnetization measurement. Jian Wei also thanks Egor Babaev, Dario Daghero, and Simon J. Bending for helpful email correspondences. Work at Peking University is supported by National Basic Research Program of China (973 Program) through Grant No. 2012CB927400 and National Natural Science Foundation of China [NSFC, Grant No. 11474008]. The work at Tulane is supported by the U.S. Department of Energy under EPSCoR Grant No. DE-SC0012432 with additional support from the Louisiana Board of Regents (support for material synthesis). The work done at SJTU is supported by by MOST of China (2012CB927403) and Penn State by DOE under DE-FG02-04ER46159.

\section*{Author contributions}
H. Wang, J.-W. Luo, and W.-J. Lou performed the experiment, J.-E. Ortmann and Z.-Q. Mao provided samples, H. Wang, J.-W. Luo, J. Wei, Z.-Q. Mao, and Y. Liu analysed data, J. Wei wrote the paper with inputs from all authors.


%

\end{document}